\documentstyle[11pt,epsf]{article}
\newtheorem{theorem}{Theorem}

\newcommand {\dfn} {\stackrel{\Delta} {=}}
\newcommand {\exe} {\stackrel{\cdot} {=}}
\newcommand {\lexe} {\stackrel{\cdot} {\le}}

\newcommand{\prm}{m^\prime}
\newcommand{\tm}{\tilde{m}}
\newcommand{\hH}{\hat{H}}
\newcommand{\hI}{\hat{I}}

\newcommand {\bx} {\mbox{\boldmath $x$}}
\newcommand {\by} {\mbox{\boldmath $y$}}

\newcommand {\bE} {\mbox{\boldmath $E$}}

\newcommand {\bX} {\mbox{\boldmath $X$}}

\newcommand{\calC}{{\cal C}}

\newcommand{\calE}{{\cal E}}
\newcommand{\calF}{{\cal F}}
\newcommand{\calG}{{\cal G}}

\newcommand{\calI}{{\cal I}}

\newcommand{\hP}{\hat{P}}
\newcommand{\calQ}{{\cal Q}}

\newcommand{\calS}{{\cal S}}
\newcommand{\calT}{{\cal T}}

\newcommand{\calX}{{\cal X}}
\newcommand{\calY}{{\cal Y}}

\begin{document}
\thispagestyle{empty}
\title{Error Exponents of Typical Random Codes}
\author{Neri Merhav}
\date{}
\maketitle

\begin{center}
The Andrew \& Erna Viterbi Faculty of Electrical Engineering\\
Technion - Israel Institute of Technology \\
Technion City, Haifa 32000, ISRAEL \\
E--mail: {\tt merhav@ee.technion.ac.il}\\
\end{center}
\vspace{1.5\baselineskip}
\setlength{\baselineskip}{1.5\baselineskip}

\begin{abstract}
We define the error exponent of the typical random code as the long--block limit of the
negative normalized {\it expectation of the logarithm} of the error probability of the random code,
as opposed to the traditional random coding error exponent, which is
the limit of the negative 
normalized {\it logarithm of the expectation} of the error probability.
For the ensemble of uniformly randomly drawn fixed composition codes,
we provide exact error exponents of typical random codes for a general discrete memoryless
channel (DMC) and a wide class of (stochastic) decoders, collectively referred to 
as the generalized likelihood decoder (GLD). This ensemble of
fixed composition codes is shown to be no worse than any other ensemble of independent
codewords that are drawn under a permutation--invariant distribution (e.g., i.i.d.\
codewords). We also present relationships between the error
exponent of the typical random code and the ordinary random coding error exponent, as well as the expurgated
exponent for the GLD. Finally, we demonstrate that our analysis technique is
applicable also to more general communication scenarios, such as list decoding (for
fixed--size lists) as well as decoding with an erasure/list 
option in Forney's sense.\\

\end{abstract}

\newpage
\section{Introduction}
\label{intro}

Traditionally, the random coding error exponent is defined as
\begin{equation}
\label{rce}
E_{\mbox{\tiny r}}(R)=\lim_{n\to\infty}\left[-\frac{\ln\bE P_{\mbox{\tiny
e}}(\calC_n)}{n}\right], 
\end{equation}
where $n$ is the
block length, $R$ is the coding rate,
$P_{\mbox{\tiny e}}(\calC_n)$ is the error probability of a codebook
$\calC_n$, and the
expectation is with respect to (w.r.t) the randomness of $\calC_n$ across the ensemble of codes
(see, e.g., \cite{CK11}, \cite{Gallager68}, \cite{VO79} and many references
therein). While fairly
easy to analyze (or at least, to bound), the random coding error exponent is
also known to be quite a
pessimistic performance measure because, at low coding
rates, $\bE P_{\mbox{\tiny e}}(\calC_n)$ is dominated by relatively poor codes
in the ensemble, rather than by the channel noise. Indeed, at low coding
rates, the random coding bound can be improved by the well known expurgation
idea \cite{CK11}, \cite{Gallager68}, \cite{VO79}.

An alternative ensemble performance metric, that is never worse 
than the random coding error exponent,
and in fact, strictly better at low rates, is the error exponent of the 
{\it typical random code} (TRC), which we define
by simply commuting the expectation operator with the logarithmic function in
(\ref{rce}), i.e.,
\begin{equation}
E_{\mbox{\tiny trc}}(R)=\lim_{n\to\infty}\left[-\frac{\bE\ln
P_{\mbox{\tiny e}}(\calC_n)}{n}\right], 
\end{equation}
provided that the limit exists. The fact that $E_{\mbox{\tiny trc}}(R)$ cannot
be smaller than $E_{\mbox{\tiny r}}(R)$ is easily understood from Jensen's
inequality, but to capture the insight behind the different meanings of these two
exponents, consider the following informal, intuitive consideration: let
$\calS(E)$ be the
collection of all codes $\{\calC_n\}$ in the ensemble,
with $P_{\mbox{\tiny e}}(\calC_n)\approx e^{-nE}$ for
a given value of $E$. Then, 
$\bE P_{\mbox{\tiny e}}(\calC_n)\approx\sum_E
\mbox{Pr}\{\calS(E)\}\cdot e^{-nE}$ (approximating by a discrete grid of values of $E$, for
simplicity), a quantity
that is dominated by the codes in
$\calS(E^*)$, where $E^*$ maximizes the product $\mbox{Pr}\{\calS(E)\}\cdot e^{-nE}$.
The codes of $\calS(E^*)$ are the ``poor'' 
codes that we have referred to in the previous paragraph, and $E_{\mbox{\tiny
r}}(R)$ is given by $E^*$ plus the exponential rate of
$\mbox{Pr}\{\calS(E^*)\}$.
On the other hand, $E_{\mbox{\tiny trc}}(R)$ is approximately
equal to $\sum_E\mbox{Pr}\{\calS(E)\}\cdot E$, 
and if there is one value of $E$, say $E_0$, at which
$\mbox{Pr}\{\calS(E)\}$ concentrates in the large $n$ limit, 
then the members of $\calS(E_0)$ are the typical
codes for our purpose, and $E_{\mbox{\tiny trc}}(R)=E_0$. We will see
later on that indeed, such a concentration property takes place, and hence the
notion of ``typical random codes''. 
Generally speaking, we believe that the TRC
error exponent should be
the more relevant quantity of interest, because the code is selected randomly
once and for all, and then it is natural to ask what would be the error
exponent associated with the typical code.\footnote{Interestingly, there is an analogous
consideration in statistical mechanics of disordered systems, which are modeled with random
parameters. According to these models, Nature ``selects'' those random parameters just once, and
one is interested in the free energy of a typical realization of the system,
which is given in terms of the expectation of the logarithm of partition function
(a.k.a.\ quenched
average), rather than the logarithm of the expectation of the partition function
(annealed average), see, e.g., \cite[Sect.\ 5.7]{sp+it}.}

The problem is that
it is considerably more difficult to analyze the expectation of the logarithm of the error
probability than the logarithm of the expected error probability. 
This is true especially if one insists on obtaining exact error exponents and not just
bounds. Perhaps this is one of the main reasons that not much earlier work has
been done on error exponents of TRC's. The most relevant exception to this
rule is the brief article by Barg and Forney \cite{BF02} (see also
\cite{Forney05}), where among other
things, they have derived the error exponent of the TRC for the binary symmetric
channel (BSC) w.r.t.\ the ensemble of codes drawn by fair coin tossing of each
bit of each codeword. In \cite{BF02}, Barg and Forney have shown that
at a certain range of low rates, $E_{\mbox{\tiny trc}}(R)$ lies between
$E_{\mbox{\tiny r}}(R)$ and the expurgated exponent, 
$E_{\mbox{\tiny ex}}(R)$, and there is an interesting relationship between
$E_{\mbox{\tiny trc}}(R)$ and the expurgated exponent function $E_{\mbox{\tiny
ex}}(\cdot)$ (also applicable to low rates), given by
\begin{equation}
\label{trc-ex relation}
E_{\mbox{\tiny trc}}(R)=E_{\mbox{\tiny ex}}(2R)+R.
\end{equation}
Other related works, with some linkage to error exponents of TRC's, can be found
in the statistical physics literature \cite{Kabashima08}, \cite{MR06},
\cite{SMKS03}, where the replica method and the cavity method have been largely used mostly in the
context of low--density parity--check (LDPC) codes.

In this work, we propose a systematic derivation of exact error exponents of
TRC's. This extends the corresponding results of \cite{BF02} in several
directions.
\begin{enumerate}
\item A general DMC is considered, not merely the BSC.
\item The analysis covers a wide family of stochastic decoders, not only the maximum
likelihood (ML) decoder (but the ML decoder is a special case).
\item We adopt the ensemble of constant composition codes, with independent codewords
drawn under the uniform distribution across a given type class. This random
coding distribution 
is shown to be no worse than any other permutation--invariant distribution,
including, of course, the i.i.d.\ distribution, as in \cite{BF02}.
\item It is shown that the relation (\ref{trc-ex relation}) continues to hold even
in the more general scenario, as described in items 1--3 above. Moreover,
using the improved expurgated exponent of \cite{gld}, it is shown that
eq.\ (\ref{trc-ex relation}) holds for the entire range of rates, not merely at
low rates.
\item It is demonstrated that the proposed analysis technique of TRC error exponents 
is applicable also to more
general scenarios, such as list decoding (with fixed list size) as well as
decoding with an erasure/list option in Forney's sense \cite{Forney68}.
\end{enumerate}

It should be pointed out that in \cite[p.\ 2572, right column, comment no.\ 6]{BF02}, 
Barg and Forney comment that it is possible to
extend the derivation to general DMC's with the ensemble of constant composition codes, but
they have not displayed this extension, and it is not trivial
to guess, from their analysis for the BSC and i.i.d.\ random coding,
what is the TRC error exponent formula for a general DMC under the ensemble of
constant composition codes and the more general decoders that we consider here.

The starting point of our analysis approach is similar to that of the well
known replica method, a popular technique borrowed from statistical physics 
(see, e.g., \cite[Sect.\ 4.5]{sp+it} and references therein), 
but this is the only point of similarity between our
method and the replica method. In particular, it is 
based on the identity
\begin{equation}
\label{replicaidea}
\bE\ln P_{\mbox{\tiny e}}(\calC_n)=\lim_{\rho\to\infty}\ln\left(\bE
[P_{\mbox{\tiny e}}(\calC_n)]^{1/\rho}\right)^\rho=
\lim_{\rho\to\infty}\rho\ln\left(\bE
[P_{\mbox{\tiny e}}(\calC_n)]^{1/\rho}\right).
\end{equation}
This ingredient of calculating the ($1/\rho$)--th moment of the probability of
error and raising it to the power of $\rho$, is also the
technique used in the derivation 
of expurgated exponents \cite{Gallager68}, \cite{VO79}, the only
difference is that in the context of expurgated bounds, 
this is applied to $P_{\mbox{\tiny e}|m}(\calC_n)$, 
the conditional error probability given that message $m$ was transmitted, and
then an expurgation argument is applied to assert that upon eliminating bad
codewords from the code, we end up with a code for which $P_{\mbox{\tiny
e}|m}(\calC_n)$ is upper bounded by a certain quantity, for all remaining
messages. Here,
on the other hand, we wish to invoke (\ref{replicaidea}) for the overall error
probability, 
\begin{equation}
\label{overallpe}
P_{\mbox{\tiny e}}(\calC_n)=\frac{1}{M}\sum_{m=0}^{M-1}P_{\mbox{\tiny
e}|m}(\calC_n),
\end{equation}
where $M$ is the number of codebook messages.
Nonetheless, this difference between the two derivations 
is not dramatic, and it is therefore not too
surprising that the TRC error exponent and the expurgated exponent are
related.

The outline of the remaining part of the paper is as follows.
In Section 2, we establish notation conventions.
In Section 3, we describe the setup, provide formal definitions, and spell out
the objectives. In Section 4, we present the main result, a single--letter
formula for the TRC error exponent, and discuss some of its properties.
In Section 5, we prove the main result, and finally, in Section 6, we
demonstrate how the same technique can be used to derive TRC error exponents
in more general scenarios.

\section{Notation Conventions}
\label{notation}

Throughout the paper, random variables will be denoted by capital
letters, specific values they may take will be denoted by the
corresponding lower case letters, and their alphabets
will be denoted by calligraphic letters. Random
vectors and their realizations will be denoted,
respectively, by capital letters and the corresponding lower case letters,
both in the bold face font. Their alphabets will be superscripted by their
dimensions. For example, the random vector $\bX=(X_1,\ldots,X_n)$, ($n$ --
positive integer) may take a specific vector value $\bx=(x_1,\ldots,x_n)$
in $\calX^n$, the $n$--th order Cartesian power of $\calX$, which is
the alphabet of each component of this vector.
Sources and channels will be denoted by the letters $P$, $Q$ and $W$,
subscripted by the names of the relevant random variables/vectors and their
conditionings, if applicable, following the standard notation conventions,
e.g., $Q_X$, $P_Y$, $W_{Y|X}$, and so on. When there is no room for ambiguity, these
subscripts will be omitted.
The probability of an event $\calG$ will be denoted by $\mbox{Pr}\{\calG\}$,
and the expectation
operator with respect to (w.r.t.) a probability distribution $P$ will be
denoted by
$\bE_P\{\cdot\}$. Again, the subscript will be omitted if the underlying
probability distribution is clear from the context.
For two
positive sequences $a_n$ and $b_n$, the notation $a_n\exe b_n$ will
stand for equality in the exponential scale, that is,
$\lim_{n\to\infty}\frac{1}{n}\log \frac{a_n}{b_n}=0$. Similarly,
$a_n\lexe b_n$ means that
$\limsup_{n\to\infty}\frac{1}{n}\log \frac{a_n}{b_n}\le 0$, and so on.
The indicator function
of an event $\calG$ will be denoted by $\calI\{\calG\}$. The notation $[x]_+$
will stand for $\max\{0,x\}$.

The empirical distribution of a sequence $\bx\in\calX^n$, which will be
denoted by $\hat{P}_{\bx}$, is the vector of relative frequencies
$\hat{P}_{\bx}(x)$
of each symbol $x\in\calX$ in $\bx$.
The type class of $\bx\in\calX^n$, denoted $\calT(\hP_{\bx})$, is the set of
all
vectors $\bx'$
with $\hat{P}_{\bx'}=\hat{P}_{\bx}$.
Information measures associated with empirical distributions
will be denoted with `hats' and will be subscripted by the sequences from
which they are induced. For example, the entropy associated with
$\hat{P}_{\bx}$, which is the empirical entropy of $\bx$, will be denoted by
$\hat{H}_{\bx}(X)$.
Similar conventions will apply to the joint empirical
distribution, the joint type class, the conditional empirical distributions
and the conditional type classes associated with pairs (and multiples) of
sequences of length $n$.
Accordingly, $\hP_{\bx\by}$ will be the joint empirical
distribution of $(\bx,\by)=\{(x_i,y_i)\}_{i=1}^n$,
and $\calT(\hP_{\bx\by})$ will denote
the joint type class of $(\bx,\by)$. Similarly, $\calT(\hP_{\bx|\by}|\by)$
will stand for
the conditional type class of $\bx$ given
$\by$, $\hH_{\bx\by}(X,Y)$
will designate the empirical joint entropy of $\bx$
and $\by$,
$\hH_{\bx\by}(X|Y)$ will be the empirical conditional entropy,
$\hI_{\bx\by}(X;Y)$ will
denote empirical mutual information, and so on.
We will also use similar rules of notation in the context of
a generic distribution, $Q_{XY}$ (or $Q$, for short, when there is no risk of
ambiguity): we use
$\calT(Q_X)$ for the type class of sequences with empirical distribution
$Q_X$,
$H_Q(X)$ -- for the corresponding empirical entropy,
$\calT(Q_{XY})$ -- for the joint type class,
$T(Q_{X|Y}|\by)$ -- for the conditional type class of $\bx$ given $\by$,
$H_Q(X,Y)$ -- for the joint empirical entropy,
$H_Q(X|Y)$ -- for the conditional empirical entropy of $X$ given $Y$,
$I_Q(X;Y)$ -- for the empirical mutual information, and so on.
We will also use the customary notation for the weighted divergence,
\begin{equation}
D(Q_{Y|X}\|P_{Y|X}|Q_X)=\sum_{x\in\calX}Q_X(x)\sum_{y\in\calY}Q_{Y|X}(y|x)\log
\frac{Q_{Y|X}(y|x)}{P_{Y|X}(y|x)}.
\end{equation}
Finally, the notation $(Q_Y\odot Q_{X|Y})_X$ will stand for the $X$--marginal induced by $Q_Y$ and
$Q_{X|Y}$, that is, $(Q_Y\odot Q_{X|Y})_X(x)=\sum_yQ_Y(y)Q_{X|Y}(x|y)$.

\section{Formulation, Definitions, and Main Result}
\label{formulation}

Consider a DMC, $W=\{W(y|x),~x\in\calX,~y\in\calY\}$, where $\calX$ is a
finite input alphabet, $\calY$ is a finite output alphabet, and $W(y|x)$ is the
channel input--output single--letter transition probability from $x$ to $y$. When  fed by a
vector $\bx=(x_1,x_2,\ldots,x_n)\in\calX^n$, the channel responds by producing an output
vector $\by=(y_1,y_2,\ldots,y_n)\in\calY^n$, according to
\begin{equation}
W(\by|\bx)=\prod_{i=1} W(y_i|x_i).
\end{equation}

Let $\calC_n=\{\bx_0,\bx_1,\ldots,\bx_{M-1}\}\subseteq\calX^n$, $M=e^{nR}$, $R$ being the
coding rate in nats per channel use. When the transmitter wishes to convey 
a message $m\in\{0,1,\ldots,M-1\}$, it feeds the channel with $\bx_m$. We
consider the ensemble of fixed composition codes, where each codeword is
selected independently at random under the uniform distribution across a
given type class of $n$--vectors, $\calT(Q_X)$.

As in \cite{gld} and \cite{p199}, we consider a generalized 
version of the so called {\it likelihood decoder}
\cite{SMF15}, \cite{SCP14}, \cite{YAG13}, which is a stochastic decoder that
randomly selects the message estimate according to the posterior
probability distribution given $\by$. The generalized likelihood decoder (GLD) considered
here, randomly selects the decoded message according to the generalized
posterior, 
\begin{equation}
P(\hat{m}=m|\by)=\frac{\exp\{ng(\hP_{\bx_m\by})\}}{\sum_{m^\prime=0}^{M-1}
\exp\{ng(\hP_{\bx_{m^\prime}\by})\}},
\end{equation}
where $g(\cdot)$, henceforth
referred to as the {\it decoding metric}, 
is an arbitrary continuous functional of a joint distribution
$Q_{XY}$ on $\calX\times\calY$. For
\begin{equation}
\label{ordinary1}
g(Q_{XY})=\sum_{x\in\calX}\sum_{y\in\calY}Q_{XY}(x,y)\ln
W(y|x),
\end{equation}
we recover the ordinary likelihood decoder as in \cite{SMF15},
\cite{SCP14}, \cite{YAG13}. For
\begin{equation}
\label{ordinary2}
g(Q_{XY})=\beta\sum_{x\in\calX}\sum_{y\in\calY}Q_{XY}(x,y)\ln
W(y|x),
\end{equation}
$\beta\ge 0$ being a free parameter, we
extend this to a parametric family of decoders, where
$\beta$ controls the skewedness of the posterior. In particular,
$\beta\to\infty$ leads to the (deterministic)  ML decoder.
Other interesting choices are associated with mismatched metrics,
\begin{equation}
g(Q_{XY})=\beta\sum_{x\in\calX}\sum_{y\in\calY}Q_{XY}(x,y)\ln
W^\prime(y|x),
\label{mismatched}
\end{equation}
$W^{\prime}$ being different from $W$, and
\begin{equation}
\label{mmi}
g(Q_{XY})=\beta I_Q(X;Y),
\end{equation}
which for $\beta\to\infty$,
approaches the well known universal maximum mutual information (MMI) decoder 
\cite{CK11} (see also discussion around
eqs.\ (5)--(7) of \cite{gld}).
The probability of error, associated with a given code $\calC_n$ and the GLD,
is given by
\begin{equation}
\label{explicit-pe}
P_{\mbox{\tiny e}}(\calC_n)=\frac{1}{M}\sum_{m=0}^{M-1}\sum_{m^\prime\ne
m}\sum_{\by\in\calY^n}W(\by|\bx_m)\cdot
\frac{\exp\{ng(\hP_{\bx_m\by})\}}{\sum_{\tilde{m}=0}^{M-1}
\exp\{ng(\hP_{\bx_{\tilde{m}}\by})\}}.
\end{equation}
For the ensemble of rate--$R$ fixed composition codes of type $Q_X$,
we define the random coding error exponent w.r.t.\ ML decoding (i.e., with $g$ 
given by (\ref{ordinary2}) at the limit $\beta\to\infty$), by
\begin{equation}
\label{rce-def}
E_{\mbox{\tiny r}}(R,Q_X)=\lim_{n\to\infty}\left[-\frac{\ln
[\bE P_{\mbox{\tiny e}}(\calC_n)]}{n}\right],
\end{equation}
as well as the TRC error exponent, associated with the decoding metric $g$,
\begin{equation}
\label{trc-def}
E_{\mbox{\tiny trc}}^g(R,Q_X)=\lim_{n\to\infty}\left[-\frac{\bE\ln
[P_{\mbox{\tiny e}}(\calC_n)]}{n}\right],
\end{equation}
provided that the limits exist,\footnote{The limit is well known to exist for
(\ref{rce-def}). As for (\ref{trc-def}), it will be evident from the analysis.}
and where the expectation is w.r.t.\ the randomness of $\calC_n$.
The TRC error exponent associated with the ML decoder
will be denoted by
$E_{\mbox{\tiny trc}}(R,Q_X)$, and the one for the stochastic MMI decoder
(that is, (\ref{mmi}) with $\beta=1$) will be denoted by $E_{\mbox{\tiny
trc}}^{\mbox{\tiny smmi}}(R,Q_X)$.
The main objective of this paper is to derive an exact single--letter formula
for $E_{\mbox{\tiny trc}}^g(R,Q_X)$ and to study some of its properties.

\section{Main Result}
\label{main}

Before we present the main result, we need two more few definitions:
\begin{equation}
\label{alpha-def}
\alpha(R,Q_Y)\dfn\sup_{\{Q_{X|Y}:~I_Q(X;Y)\le R,~(Q_Y\odot
Q_{X|Y})_X=Q_X\}}[g(Q_{XY})-I_Q(X;Y)]+R,
\end{equation}
and
\begin{eqnarray}
\label{Gamma-def}
\Gamma(Q_{XX^\prime},R)&\dfn&\inf_{Q_{Y|XX^\prime}}\{D(Q_{Y|X}\|W|Q_X)+I_Q(X^\prime;Y|X)+\nonumber\\
& &[\max\{g(Q_{XY}),\alpha(R,Q_Y)\}-g(Q_{X^\prime Y})]_+\}.
\end{eqnarray}
Our main result is the following theorem, whose proof appears in Section
\ref{proof}.
\begin{theorem}
Consider the setting described in Section \ref{formulation}. Then,
\begin{equation}
\label{trcee}
E_{\mbox{\tiny trc}}^g(R,Q_X)=\inf_{\{Q_{X^\prime|X}:~I_Q(X;X^\prime)\le
2R,~Q_{X^\prime}=Q_X\}}\{\Gamma(Q_{XX^\prime},R)+
I_Q(X;X^\prime)-R\}.
\end{equation}
\end{theorem}
The remaining part of this section is devoted to a discussion on Theorem 1 and
its implications.\\

\noindent
{\bf Relation to the random coding error exponent.}
In principle, the random coding error exponent is obtained by setting $\rho=1$ 
in the r.h.s.\ of (\ref{replicaidea}) instead of taking the limit $\rho\to\infty$.
We first show directly that $E_{\mbox{\tiny trc}}(R,Q_X)$ indeed cannot be
smaller than $E_{\mbox{\tiny r}}(R,Q_X)$ at any rate $R$. 
Beyond the fact that this is a good sanity check, it is insightful to
identify the origins of possible gaps between the two exponents.
To this end, let us
examine $E_{\mbox{\tiny trc}}^{\mbox{\tiny smmi}}(R,Q_X)$, that is, as
mentioned before, defined
for the sub-optimal GLD based on $g(Q)=I_Q(X;Y)$ (and which
is especially convenient to work with). In this case, it can be readily
verified that $\alpha(R,Q_Y)=R$, which yields
\begin{equation}
\Gamma(Q_{XX^\prime})=\min_{Q_{Y|XX^\prime}}\{D(Q_{Y|X}\|W|Q_X)+I_Q(X^\prime;Y|X)+
[\max\{I_Q(X;Y),R\}-I_Q(X^\prime;Y)]_+\},
\end{equation}
and so,
\begin{eqnarray}
E_{\mbox{\tiny trc}}(R,Q_X)
&\ge&E_{\mbox{\tiny trc}}^{\mbox{\tiny smmi}}(R,Q_X)\nonumber\\
&=&\min_{\{Q_{X^\prime Y|X}:~I_Q(X;X^\prime)\le 2R,~Q_{X^\prime}=Q_X\}}\left\{
D(Q_{Y|X}\|W|Q_X)+I_Q(X^\prime;Y|X)+I_Q(X;X^\prime)+\right.\nonumber\\
& &\left.[\max\{I_Q(X;Y),R\}-I_Q(X^\prime;Y)]_+-R\right\}\nonumber\\
&=&\min_{\{Q_{X^\prime Y|X}:~I_Q(X;X^\prime)\le 2R,~Q_{X^\prime}=Q_X\}}\left\{
D(Q_{Y|X}\|W|Q_X)+I_Q(X^\prime;X|Y)+I_Q(X^\prime;Y)+\right.\nonumber\\
& &\left.[\max\{I_Q(X;Y),R\}-I_Q(X^\prime;Y)]_+-R\right\}\nonumber\\
&=&\min_{\{Q_{X^\prime Y|X}:~I_Q(X;X^\prime)\le 2R,~Q_{X^\prime}=Q_X\}}\left\{
D(Q_{Y|X}\|W|Q_X)+I_Q(X^\prime;X|Y)+\right.\nonumber\\
& &\left.\max\{I_Q(X;Y),I_Q(X^\prime;Y),R\}-R\right\}\nonumber\\
&=&\min_{\{Q_{X^\prime Y|X}:~I_Q(X;X^\prime)\le 2R,~Q_{X^\prime}=Q_X\}}\left\{
D(Q_{Y|X}\|W|Q_X)+I_Q(X^\prime;X|Y)+\right.\nonumber\\
& &\left.[\max\{I_Q(X;Y),I_Q(X^\prime;Y)\}-R]_+\right\}\nonumber\\
&\ge&\min_{Q_{Y|X}}\left\{
D(Q_{Y|X}\|W|Q_X)+
[I_Q(X;Y)-R]_+\right\}\nonumber\\
&=&E_{\mbox{\tiny r}}(R,Q_X),
\end{eqnarray}
where the first inequality is because the metric $g(Q)=I_Q(X;Y)$ may be
sub-optimal and the second inequality is because we have dropped the
constraints and the terms $I_Q(X^\prime;X|Y)$, 
and $I_Q(X^\prime;Y)$. The last equality is well known (see, e.g.,
\cite{CK11}). 

\noindent
{\bf Relation to the expurgated exponent.}
It should be pointed out that in \cite{gld},
the following expurgated bound was found for random fixed composition codes
and the GLD:
\begin{equation}
\label{expurgated}
E_{\mbox{\tiny ex}}^g(R,Q_X)=\inf_{\{Q_{XX^\prime}:~I_Q(X;X^\prime)\le
R,~Q_{X^\prime}=Q_X\}}\{\Gamma(Q_{XX^\prime},R)+I_Q(X;X^\prime)-R\},
\end{equation}
and it has been shown in \cite{gld} that for ML decoding, this expurgated
exponent is at least
as large as the Csisz\'ar--K\"orner--Marton (CKM) expurgated exponent
\cite[p.\ 165, Problem 10.18]{CK11}. Obviously, we have the following simple relationship
between $E_{\mbox{\tiny trc}}^g(R,Q_X)$ and $E_{\mbox{\tiny ex}}^g(R,Q_X)$:
\begin{equation}
\label{trc-ex-gen}
E_{\mbox{\tiny trc}}^g(R,Q_X)=E_{\mbox{\tiny ex}}^g(2R,Q_X)+R,
\end{equation}
which extends the relation (\ref{trc-ex relation}) quite considerably.
This relation is understood from the following consideration: as mentioned in
the Introduction, the difference
between the TRC error exponent and the expurgated
exponent is that the former is applied to the overall error probability
(\ref{overallpe}), whereas the latter is applied to the conditional error
probability given that a particular message $m$ was sent. The overall error
probability (\ref{explicit-pe}) contains a double summation over the messages,
indexed by $m$ and $m^\prime$, whose exponential rate is $2R$, as opposed to the
conditional error probability, which includes only a single
summation over $m^\prime$, whose rate is $R$, hence the argument of $2R$ in
the r.h.s.\ of (\ref{trc-ex-gen}). On the other hand, (\ref{explicit-pe})
contains normalization by $M$, which is absent in the conditional error
probability, hence the addition of $R$ on the r.h.s.\ of (\ref{trc-ex-gen}).
Clearly, for any $R$,
$E_{\mbox{\tiny
trc}}^g(R,Q_X)\le E_{\mbox{\tiny ex}}^g(R)$, as the two functions
are given by minimization of the same objective, but in $E_{\mbox{\tiny
trc}}^g(R,Q_X)$, the minimization is over a larger set of distributions.
At zero--rate, we have
$E_{\mbox{\tiny trc}}^g(0,Q_X)=E_{\mbox{\tiny ex}}^g
(0,Q_X)$, which for ML decoding,
is strictly larger than $E_{\mbox{\tiny r}}(0,Q_X)$, in general.
From continuity, it appears then that there is at least some range of
low rates where the TRC error exponent is strictly larger than the random
coding error exponent, but above a certain rate, the two
exponents may coincide.

\noindent
{\bf ML decoding.}
An important special case is, of course, the optimal ML decoder, which as
mentioned earlier,
corresponds to the choice $g(Q)=\beta E_Q\ln W(Y|X)$ for $\beta\to\infty$.
For very large $\beta$, $\alpha(R,Q_Y)\approx\beta a(R,Q_Y)$, where
\begin{equation}
a(R,Q_Y)=\sup_{\{Q_{X|Y}:~I_Q(X;Y)\le R,~(Q_Y\odot Q_{X|Y})_X=Q_X\}}\bE_Q\ln W(Y|X).
\end{equation}
As $\beta\to\infty$, the term $\beta[\max\{E_Q\ln W(Y|X),a(R,Q_Y)\}-E_Q\ln
W(Y|X^\prime)]_+$, that appears in the objective, disappears, and instead,
there is an additional constraint
that the expression in the square brackets of that term, would vanish. In other words,
the result is
\begin{equation}
E_{\mbox{\tiny trc}}(R,Q_X)=
\inf_{Q_{X^\prime
Y|X}\in\calS(R)}
\{D(Q_{Y|X}\|W|Q_X)+I_Q(X^\prime;X,Y)\}-R,
\end{equation}
where
\begin{eqnarray}
\calS(R)&\dfn&\{Q_{X^\prime Y|X}:~I_Q(X;X^\prime)\le
2R,~Q_{X^\prime}=Q_X,\nonumber\\
& &E_Q\ln
W(Y|X^\prime)\ge\max\{E_Q\ln
W(Y|X),a(R,Q_Y)\}\}.
\end{eqnarray}
It is interesting to note that the third constraint in 
$\calS(R)$ designates the event that an
incorrect codeword (represented by $X^\prime$) 
receives a log--likelihood score higher than that of the correct codeword (represented by
$X$) as well as those of all other
codewords (represented by the
term $a(R,Q)$).
The term $a(R,Q)$ designates the typical value (with an extremely
high probability) of the highest log--likelihood score among all the remaining
incorrect codewords.\footnote{Observe that $a(R,Q)$ can be interpreted as the
negative distortion--rate function (the inverse of the rate--distortion
function) of a ``source'' $Q_Y$ w.r.t.\ the distortion
measure $d(x,y)=-\ln W(y|x)$ and the additional constraint that
the ``output'' distribution would be $Q_X$.} To understand the intuition behind this interpretation,
observe that given a channel output $\by\in\calT(Q_Y)$, the probability that a
randomly chosen codeword from $\calT(Q_X)$ would fall in a given conditional type,
$\calT(Q_{X|Y}|\by)$, is of the exponential order of $e^{-nI_Q(X;Y)}$. Therefore, if we select
$e^{nR}$ codewords at random, all conditional types with $I_Q(X;Y) < R$ will be
populated with very high probability. Among these conditional types, the
highest log--likelihood score would be $\sup_{\{Q_{X|Y}:~I_Q(X;Y)\le
R,~(Q_Y\odot Q_{X|Y})_X=Q_X\}}\bE_Q\ln
W(Y|X)$, which is exactly $a(R,Q_Y)$. This 
replaces the traditional
union of {\it pairwise} error events, by the union of {\it disjoint} error events,
where in each one of them, one incorrect 
codeword receives a score higher than all the others (not just higher than
that of the correct
codeword alone). As these events are disjoint, the
probability of their union is equal to the sum of probabilities, i.e., the union bound is
tight in this case.

\noindent
{\bf Other ensembles with permutation--invariant random coding distributions.}
So far we considered only the ensemble of fixed composition codes, namely,
each codeword was selected independently at random under the uniform
distribution within $\calT(Q_X)$.
Consider, more generally, a probability distribution over $\calX^n$ with the following two
properties:
\begin{enumerate}
\item If $\bx$ and $\bx^\prime$ belong to the same type, then
$P(\bx)=P(\bx^\prime)$. In other words, the distribution is uniform within
each type.
\item There exists a function, $\Delta(Q_X)\ge 0$, such that for every
sequence, $\{Q_X^n\}$,
of rational distributions with denominator $n$,
and every $Q_X$ at which $\Delta(Q_X)$ is continuous,
$Q_X^n\to Q_X$ implies $\lim_{n\to\infty}[-\frac{1}{n}\log
P\{\calT(Q_X^n)\}]=\Delta(Q_X)$.
\end{enumerate}
For example, if $P$ is i.i.d., $\Delta(Q_X)=D(Q_X\|P)$. The ensemble of fixed
composition codes also satisfies these requirements, provided that we allow some
small tolerance $\delta$ in the empirical distribution rather than insisting
on an
exact empirical distribution,\footnote{This small modification does not have
any essential impact on the results.} and
then
\begin{equation}
\Delta(Q_X)=\left\{\begin{array}{ll}
0 & d(Q_X,Q_X^*)\le\delta\\
\infty & d(Q_X,Q_X^*)>\delta\end{array}\right.
\end{equation}
where $d(\cdot,\cdot)$ is some distance measure in the space of distributions
over $\calX$.

It turns out, however, that there is nothing really to gain from this extension in terms of
performance. In other words, among all ensembles of this family, the
one of fixed composition codes, that we have studied thus far, is essentially
the best. To see why this is true, consider the following argument, which is
largely quite standard.
Given a code $\calC_n$, let $\calC_n(Q_X)\dfn\calC_n\cap\calT(Q_X)$,
$M(Q_X)\dfn|\calC_n(Q_X)|$ and
$R(Q_X)\dfn\frac{1}{n}\log M(Q_X)$. 
Obviously, there must be at least one $Q_X$ 
for which $\Delta(Q_X)=0$,
since the number of
different types is sub--exponential in $n$. 
Let us denote one of the distributions with this property by $Q_X^*$.
Now, for every given $\calC_n$, we have
\begin{eqnarray}
P_{\mbox{\tiny e}}(\calC_n)
&=&\frac{1}{M}\sum_{m=0}^{M-1}P_{\mbox{\tiny
e}|m}(\calC_n)\nonumber\\
&=&\sum_{Q_X}\frac{M(Q_X)}{M}\cdot\frac{1}{M(Q_X)}\sum_{m:~\bx_m\in\calC_n(Q_X)}P_{\mbox{\tiny
e}|m}(\calC_n)\nonumber\\
&\ge&\sum_{Q_X}\frac{M(Q_X)}{M}\cdot\frac{1}{M(Q_X)}\sum_{m:~\bx_m\in\calC_n(Q_X)}P_{\mbox{\tiny
e}|m}[\calC_n(Q_X)]\nonumber\\
&=&\sum_{Q_X}\frac{M(Q_X)}{M}P_{\mbox{\tiny e}}[\calC_n(Q_X)]\nonumber\\
&\ge&e^{n[R(Q_X)-R]}P_{\mbox{\tiny e}}[\calC_n(Q_X)],
\end{eqnarray}
where the last inequality holds for every $Q_X$, and so,
\begin{equation}
\ln P_{\mbox{\tiny e}}(\calC_n)\ge\ln P_{\mbox{\tiny
e}}[\calC_n(Q_X)]+n[R(Q_X)-R].
\end{equation}
Now, for every $\epsilon > 0$,
as long as $\Delta(Q_X) < R$, with very high probability (tending to 1
double--exponentially rapidly w.r.t.\ the new
ensemble), we will have $R(Q_X)\ge R-\Delta(Q_X)-\epsilon$, and in particular,
$R(Q_X^*)\ge R-\Delta(Q_X^*)-\epsilon = R-\epsilon$, and so,
for every such code
\begin{equation}
\ln P_{\mbox{\tiny e}}(\calC_n)\ge\ln P_{\mbox{\tiny
e}}[\calC_n(Q_X^*)]-n\epsilon.
\end{equation}
Let $\calG_n$ denote the collection of codes with $R(Q_X)\ge
R-\Delta(Q_X)-\epsilon$ for all $Q_X$ such that $\Delta(Q_X) < R$,
and observe that the probability of $\calG_n$ is overwhelmingly large for
large $n$. Then,
\begin{eqnarray}
\bE\left\{\ln P_{\mbox{\tiny e}}(\calC_n)\right\}&=&
\sum_{\calC_n}P(\calC_n)\ln P_{\mbox{\tiny e}}(\calC_n)\nonumber\\
&=&\sum_{\calC_n\in\calG_n}P(\calC_n)\ln P_{\mbox{\tiny e}}(\calC_n)+
\sum_{\calC_n\in\calG_n^c}P(\calC_n)\ln P_{\mbox{\tiny
e}}(\calC_n)\nonumber\\
&\ge&P(\calG_n)\cdot\sum_{\calC_n\in\calG_n}
P(\calC_n|\calG_n)\ln P_{\mbox{\tiny
e}}[\calC_n(Q_X^*)]-n\epsilon-n[E_{\mbox{\tiny
sp}}(R)+o(n)]P(\calG_n^c)\nonumber\\
&\ge&-n[E_{\mbox{\tiny trc}}^g
(R-\epsilon,Q_X^*)+O(\epsilon)]P(\calG_n)
-n\epsilon-n[E_{\mbox{\tiny sp}}(R)+o(n)]P(\calG_n^c),
\end{eqnarray}
where we have used the fact that, due to the uniformity of the random coding
distribution within each type, under $P(\cdot|\calG_n)$, the sub-code
$\calC_n(Q_X^*)$ is a randomly selected
fixed composition code of rate at least $R-\epsilon$.
Now, since $P(\calG_n^c)$ is double exponentially small, the right--most side
is essentially $-nE_{\mbox{\tiny trc}}^g(R,Q_X^*)$ for small
$\epsilon$.

\section{Proof of Theorem 1}
\label{proof}

The proof of Theorem 1 is divided into two parts. In the first part, we prove
that the TRC error exponent is lower bounded by the r.h.s.\ of eq.\
(\ref{trcee}). In this proof, there are a few steps (such as the inequalities
in eqs.\ (\ref{sheva}) and (\ref{enumerators}) in the sequel) where it is not obvious that
exponential tightness is not compromised, and therefore, we need 
the second part, where we prove that the TRC error exponent is also upper bounded by the same
expression. Obviously, for the former, we need an upper bound on the error
probability, whereas for the latter, we need a lower bound.

\subsection{Lower Bound on the TRC Error Exponent}

\begin{eqnarray}
P_{\mbox{\tiny e}}(\calC_n)&=&\frac{1}{M}\sum_{m=0}^{M-1}\sum_{\prm\ne
m}\sum_{\by}W(\by|\bx_m)\cdot\frac{\exp\{ng(\hP_{\bx_{\prm}\by})\}}{\sum_{\tm=0}^{M-1}
\exp\{ng(\hP_{\bx_{\tm}\by})\}}\nonumber\\
&=&\frac{1}{M}\sum_{m=0}^{M-1}\sum_{\prm\ne
m}\sum_{\by}W(\by|\bx_m)\cdot\frac{\exp\{ng(\hP_{\bx_{\prm}\by})\}}{\exp\{ng(\hP_{\bx_{m}\by})\}
+\sum_{\tm\ne m}
\exp\{ng(\hP_{\bx_{\tm}\by})\}}\nonumber\\
&\dfn&\frac{1}{M}\sum_{m=0}^{M-1}\sum_{\prm\ne
m}\sum_{\by}W(\by|\bx_m)\cdot\frac{\exp\{ng(\hP_{\bx_{\prm}\by})\}}{\exp\{ng(\hP_{\bx_{m}\by})\}
+Z_m(\by)},
\end{eqnarray}
and so, 
considering the ensemble of fixed composition codes of type $Q_X$, 
we have
\begin{eqnarray}
\bE\left\{[P_{\mbox{\tiny e}}(\calC_n)]^{1/\rho}\right\}&=&
\bE\left\{\left[\frac{1}{M}\sum_{m=0}^{M-1}\sum_{\prm\ne
m}\sum_{\by}W(\by|\bx_m)\cdot\frac{\exp\{ng(\hP_{\bx_{\prm}\by})\}}{\exp\{ng(\hP_{\bx_{m}\by})\}
+Z_m(\by)}\right]^{1/\rho}\right\}.
\end{eqnarray}
Let $\epsilon > 0$ be arbitrarily small. It is shown in 
\cite[Appendix B]{gld}, that with the possible exception of a double--exponentially
small fraction of the fixed composition codes of type $Q_X$,
all other codes in this class satisfy
\begin{equation}
\label{zbound}
Z_m(\by)\ge \exp\{n\alpha(R-\epsilon,\hP_{\by})\},~~~~~~
\forall~m\in\{0,1,\ldots,M-1\},~\by\in\calY^n.
\end{equation}
We then have,
\begin{eqnarray}
\label{sheva}
\bE\left\{[P_{\mbox{\tiny e}}(\calC_n)]^{1/\rho}\right\}&\lexe&
\bE\left(\left[\frac{1}{M}\sum_{m=0}^{M-1}\sum_{\prm\ne
m}\sum_{\by}W(\by|\bx_m)\times\right.\right.\nonumber\\
& &\left.\left.\min\left\{1,
\frac{\exp\{ng(\hP_{\bx_{\prm}\by})\}}{\exp\{ng(\hP_{\bx_{m}\by})\}
+\exp\{n\alpha(R-\epsilon,Q_Y)\}}\right\}\right]^{1/\rho}\right),
\end{eqnarray}
where we have neglected the double--exponentially small contribution of the
codes that do not satisfy (\ref{zbound}).
Now, the inner--most sum (over $\{\by\}$) 
can be easily assessed using the method of types \cite{CK11}.
Using the arbitrariness of $\epsilon$, the
result\footnote{See \cite[Section V]{gld}.} 
is that this sum is
of the exponential order of $\exp\{-n\Gamma(\hP_{\bx_m\bx_{m^\prime}},R)\}$, 
and so,
\begin{eqnarray}
\bE\left\{[P_{\mbox{\tiny e}}(\calC_n)]^{1/\rho}\right\}&\lexe&
\bE\left(\left[\frac{1}{M}\sum_{m=0}^{M-1}\sum_{\prm\ne
m}\exp\{-n\Gamma(\hP_{\bx_m\bx_{m^\prime}},R)\}\right]^{1/\rho}\right)\nonumber\\
&=&
e^{-nR/\rho}\bE\left\{\left[\sum_{Q_{XX^\prime}}N(Q_{XX^\prime})
\exp\{-n\Gamma(Q_{XX^\prime},R)\}\right]^{1/\rho}\right\}\nonumber\\
&\lexe&e^{-nR/\rho}\sum_{Q_{XX^\prime}}\bE\{[N(Q_{XX^\prime})]^{1/\rho}\}\cdot
\exp\{-n\Gamma(Q_{XX^\prime},R)/\rho\},
\end{eqnarray}
where $N(Q_{XX^\prime})$ is the number of codeword pairs
$\{(\bx_m\bx_{m^\prime})\}$, $m^\prime\ne m$, whose joint empirical distribution is
$Q_{XX^\prime}$, i.e.,
\begin{equation}
N(Q_{XX^\prime})=\sum_{m}\sum_{m^\prime\ne
m}\calI\{(\bx_m\bx_{m^\prime})\in\calT(Q_{XX^\prime})\},
\end{equation}
and the summation is over all $\{Q_{XX^\prime}\}$ whose marginals both agree with
the given composition of the code,
$Q_X$. We note that $N(Q_{XX^\prime})$ can also be expressed as
\begin{equation}
N(Q_{XX^\prime})=\sum_{m=0}^{M-1} N(Q_{X^\prime|X}|\bx_m),
\end{equation}
where $N(Q_{X^\prime|X}|\bx_m)$ is the number of $\{\bx_{m^\prime}\}$,
$m^\prime\ne m$, that fall in the conditional type
$\calT(Q_{X^\prime|X}|\bx_m)$.
Once we have an upper bound on $\bE\{[N(Q_{XX^\prime})]^{1/\rho}\}$,  
we can use it in order to bound
\begin{equation}
\label{mainbound}
\bE\ln P_{\mbox{\tiny e}}(\calC_n)\le \ln
\left\{\sum_{Q_{XX^\prime}}\lim_{\rho\to\infty}
\left(\bE\{[N(Q_{XX^\prime})]^{1/\rho}\}\right)^\rho\cdot
\exp\{-n\Gamma(Q_{XX^\prime})\}\right\}-nR.
\end{equation}
For a given $\rho > 1$, let $s\in[1,\rho]$. Then,
\begin{eqnarray}
\label{enumerators}
\bE\left\{\left[N(Q_{XX^\prime})\right]^{1/\rho}\right\}&=&
\bE\left[\sum_{m=0}^{M-1} N(Q_{X^\prime|X}|\bx_m)\right]^{1/\rho}\nonumber\\
&=&\bE\left(\left[\sum_{m=0}^{M-1}
N(Q_{X^\prime|X}|\bx_m)\right]^{1/s}\right)^{s/\rho}\nonumber\\
&\le&\bE\left(\sum_{m=0}^{M-1}
\left[N(Q_{X^\prime|X}|\bx_m)\right]^{1/s}\right)^{s/\rho}\nonumber\\
&\le&\left(\bE\left\{\sum_{m=0}^{M-1}
\left[N(Q_{X^\prime|X}|\bx_m)\right]^{1/s}\right\}\right)^{s/\rho}\nonumber\\
&=&\left(e^{nR}\bE\left\{
\left[N(Q_{X^\prime|X}|\bx_0)\right]^{1/s}\right\}\right)^{s/\rho}\nonumber\\
&=&e^{nRs/\rho}\left(\bE\left\{
\left[N(Q_{X^\prime|X}|\bx_0)\right]^{1/s}\right\}\right)^{s/\rho},
\end{eqnarray}
where the first inequality is based on the fact that $(\sum_i a_i)^t\le\sum_i a_i^t$
whenever $\{a_i\}$ are non--negative and $t\in[0,1]$, and the second
inequality follows from the concavity of the function $f(u)=u^{s/\rho}$ when
$0 < s/\rho\le 1$. Now,
for a given $\bx_0$, $N(Q_{X^\prime|X}|\bx_0)$ is a binomial random variable
with $e^{nR}-1$ trials and success rate of the exponential order of
$e^{-nI_Q(X;X^\prime)}$. Thus, 
similarly as shown in \cite[eqs.\ (6.37),
(6.38)]{sp+it},
\begin{equation}
\bE\left\{\left[N(Q_{X^\prime|X}|\bx_0)\right]^{1/s}\right\}\exe
\left\{\begin{array}{ll}
\exp\{n[R-I_Q(X;X^\prime)]/s\} & R > I_Q(X;X^\prime)\\
\exp\{n[R-I_Q(X;X^\prime)]\} & R \le I_Q(X;X^\prime)\end{array}\right.
\end{equation}
and so,
\begin{equation}
\bE\left[N(Q_{XX^\prime})\right]^{1/\rho}\lexe
\left\{\begin{array}{ll}
\exp\{n[(s+1)R-I_Q(X;X^\prime)]/\rho\} & R > I_Q(X;X^\prime)\\
\exp\{ns[2R-I_Q(X;X^\prime)]/\rho\} & R \le I_Q(X;X^\prime)\end{array}\right.
\end{equation}
which, after minimization over $s\in[1,\rho]$, becomes
\begin{eqnarray}
\bE\left[N(Q_{XX^\prime})\right]^{1/\rho}&\le&
\min_{1\le s\le\rho}\left\{\begin{array}{ll}
\exp\{n[(s+1)R-I_Q(X;X^\prime)]/\rho\} & R > I_Q(X;X^\prime)\\
\exp\{ns[2R-I_Q(X;X^\prime)]/\rho\} & I_Q(X;X^\prime)/2 \le R \le I_Q(X;X^\prime)\\
\exp\{ns[2R-I_Q(X;X^\prime)]/\rho\} & R < I_Q(X;X^\prime)/2
\end{array}\right.\nonumber\\
&=&\left\{\begin{array}{ll}
\exp\{n[2R-I_Q(X;X^\prime)]/\rho\} & R > I_Q(X;X^\prime)\\
\exp\{n[2R-I_Q(X;X^\prime)]/\rho\} & I_Q(X;X^\prime)/2 \le R \le
I_Q(X;X^\prime)\\
\exp\{n\rho[2R-I_Q(X;X^\prime)]/\rho\} & R < I_Q(X;X^\prime)/2
\end{array}\right.\nonumber\\
&=&\left\{\begin{array}{ll}
\exp\{n[2R-I_Q(X;X^\prime)]/\rho\} & R > I_Q(X;X^\prime)/2\\
\exp\{n[2R-I_Q(X;X^\prime)]\} & R < I_Q(X;X^\prime)/2
\end{array}\right.
\end{eqnarray}
and so,
\begin{equation}
\left(\bE\left[N(Q_{XX^\prime})\right]^{1/\rho}\right)^\rho\le
\left\{\begin{array}{ll}
\exp\{n[2R-I_Q(X;X^\prime)]\} & R > I_Q(X;X^\prime)/2\\
\exp\{n\rho[2R-I_Q(X;X^\prime)]\} & R < I_Q(X;X^\prime)/2
\end{array}\right.\nonumber\\
\end{equation}
which in the limit becomes
\begin{equation}
\label{2dbinomial}
\lim_{\rho\to\infty}\left(\bE\left[\sum_m N(Q_{X^\prime|X}|\bx_m)
\right]^{1/\rho}\right)^\rho\le
\left\{\begin{array}{ll}
\exp\{n[2R-I_Q(X;X^\prime)]\} & R > I_Q(X;X^\prime)/2\\
0 & R < I_Q(X;X^\prime)/2
\end{array}\right.
\end{equation}
and substituting this into (\ref{mainbound}), yields
\begin{eqnarray}
E_{\mbox{\tiny trc}}^g
(R,Q_X)&\ge&\min_{\{Q_{XX^\prime}:~I_Q(X;X^\prime)\le 2R,~Q_{X^\prime}=Q_X\}}
\{\Gamma(Q_{XX^\prime})+I_Q(X;X^\prime)-2R+R\}\nonumber\\
&=&\min_{\{Q_{XX^\prime}:~I_Q(X;X^\prime)\le 2R,~Q_{X^\prime}=Q_X\}}
\{\Gamma(Q_{XX^\prime})+I_Q(X;X^\prime)-R\}\nonumber\\
&=&E_{\mbox{\tiny ex}}^g(2R)+R,
\end{eqnarray}
completing half of the proof of Theorem 1.

\subsection{Upper Bound on the TRC Error Exponent}

The idea of the proof is show that with high probability, the randomly
selected code is such
that $Z_{mm^\prime}(\by)$ is {\it
upper} bounded by $\exp\{n[\alpha(R+2\epsilon,\hP_{\by})+\epsilon]\}$
for sufficiently many triplets $\{(m,m^\prime,\by)\}$, and so, the denominator
of the generalized posterior can be lower bounded by an expression of the same
exponential order as before.

We begin with a simple fact that will be needed later. Consider a
joint distribution, $Q_{XX^\prime}$, that satisfies
$I_Q(X;X^\prime)< 2R$, and define
$\calE(Q_{XX^\prime})=\{\calC_n:~N(Q_{XX^\prime}) <
\exp\{n[2R-I_Q(X;X^\prime)-\epsilon\}\}$.
We have to show that $\mbox{Pr}\{\calE(Q_{XX^\prime})\}$ is
small.\footnote{This fact, together with the small probability of
the event $\{\calC_n:~N(Q_{XX^\prime})>
\exp\{n[2R-I_Q(X;X^\prime)+\epsilon\}\}$ (proved similarly) means that
$N(Q_{XX^\prime})$ concentrates around its mean. Also, when $I_Q(X;X^\prime)>
2R$, we have $\mbox{Pr}\{N(Q_{XX^\prime})\ge 1\}\le\bE\{N(Q_{XX^\prime})\}\to
0$. Therefore, the 
typical codes (referring to $\calS(E_0)$ in the Introduction),
which capture most of the probability,
are characterized by $N(Q_{XX^\prime})=0$ for all $I_Q(X;X^\prime)>2R$, and
$N(Q_{XX^\prime})\exe\exp\{n[2R-I_Q(X;X^\prime)]\}$ for all $I_Q(X;X^\prime)<2R$.}
This follows from the following consideration.
\begin{eqnarray}
\mbox{Pr}\left\{\calE(Q_{XX^\prime})\right\}&=&\mbox{Pr}\left\{
N(Q_{XX^\prime})<
\exp\{n[2R-I_Q(X;X^\prime)-\epsilon\}\right\}\nonumber\\
&=&\mbox{Pr}\left\{
N(Q_{XX^\prime})<
e^{-n\epsilon}\bE\{N(Q_{XX^\prime})\}\right\}\nonumber\\
&=&\mbox{Pr}\left\{
\frac{N(Q_{XX^\prime})}
{\bE\{N(Q_{XX^\prime})\}}-1<-(1-e^{-n\epsilon})\right\}\nonumber\\
&\le&\mbox{Pr}\left\{
\left[\frac{N(Q_{XX^\prime})}
{\bE\{N(Q_{XX^\prime})\}}-1\right]^2>(1-e^{-n\epsilon})^2\right\}\nonumber\\
&\le&\frac{\mbox{Var}\{N(Q_{XX^\prime})\}}
{(1-e^{-n\epsilon})^2\bE^2\{N(Q_{XX^\prime})\}}.
\end{eqnarray}
Now, the denominator is of the exponential order of 
$\exp\{2n[2R-I_Q(X;X^\prime)]\}$.
Using the shorthand notation
$\calI(m,m^\prime)=\calI\{(\bx_m,\bx_{m^\prime})\in\calT(Q_{XX^\prime})\}$
and $p=\bE\calI(m,m^\prime)$, the
numerator is given as follows:
\begin{eqnarray}
\mbox{Var}\{N(Q_{XX^\prime})\}&=&
\bE\{N^2(Q_{XX^\prime})\}-\bE^2\{N(Q_{XX^\prime})\}\nonumber\\
&=&\sum_{m,m^\prime}\sum_{\tilde{m},\hat{m}}\bE\{\calI(m,m^\prime)
\calI(\tilde{m},\hat{m})\}-[M(M-1)p]^2\nonumber\\
&=&\sum_{m,m^\prime}\bE\{\calI^2(m,m^\prime)
\}+\sum_{(m,m^\prime)\ne(\tilde{m},\hat{m})}
\bE\{\calI(m,m^\prime)\calI(\tilde{m},\hat{m})\}-[M(M-1)p]^2\nonumber\\
&=&M(M-1)p+M(M-1)[M(M-1)-1]p^2-[M(M-1)p]^2\nonumber\\
&=&M(M-1)p[1+(M(M-1)-1)p-M(M-1)p]\nonumber\\
&=&M(M-1)p(1-p)\nonumber\\
&\exe&\exp\{n[2R-I_Q(X;X^\prime)]\}.
\end{eqnarray}
Thus,
\begin{eqnarray}
\mbox{Pr}\left\{\calE(Q_{XX^\prime})\right\}&\lexe&
\frac{\exp\{n[2R-I_Q(X;X^\prime)]\}}
{\exp\{n[4R-2I_Q(X;X^\prime)]\}}\nonumber\\
&=&\exp\{-n[2R-I_Q(X;X^\prime)]\},
\end{eqnarray}
which tends to zero since we have assumed that $I_Q(X;X^\prime) < 2R$.
Of course, if $I_Q(X;X^\prime) < 2R-\epsilon$, then
$\mbox{Pr}\left\{\calE(Q_{XX^\prime})\right\}$ decays at least as fast as
$e^{-n\epsilon}$.

Next, for a given $\epsilon > 0$, and a given joint type,
$Q_{XX^\prime Y}$, such that $I_Q(X;X^\prime) < 2R-\epsilon$,
let us define
\begin{equation}
Z_{mm^\prime}(\by)=\sum_{\tilde{m}\ne
m,m^\prime}\exp\{ng(\hP_{\bx_{\tilde{m}}\by})\},
\end{equation}
and
\begin{eqnarray}
\calG_n(Q_{XX^\prime Y})&=&\left\{\calC_n:~\sum_m\sum_{m^\prime\ne
m}\calI\{(\bx_m\bx_{m^\prime})\in\calT(Q_{XX^\prime})\}\times\right.\nonumber\\
&
&\left.\sum_{\by\in\calT(Q_{Y|XX^\prime}|\bx_m\bx_{m^\prime})}\calI\{Z_{mm^\prime}(\by)\le
e^{n[\alpha(R+2\epsilon,Q_Y)+\epsilon]}\}\ge\right.\nonumber\\
& &\left. \exp\{n[2R-I_Q(X;X^\prime)-3\epsilon/2]\}\cdot
|\calT(Q_{Y|XX^\prime}|\bx_m\bx_{m^\prime})|\right\},
\end{eqnarray}
where $(\bx_m\bx_{m^\prime})$, in the expression
$|\calT(Q_{Y|XX^\prime}|\bx_m\bx_{m^\prime})|$, should be understood as any
pair of codewords in $\calT(Q_{XX^\prime})$ (as the specific choice of them is
immaterial for the size of $\calT(Q_{Y|XX^\prime}|\bx_m\bx_{m^\prime})$). Next
define,
\begin{equation}
\calG_n=\bigcap_{\{Q_{XX^\prime Y}:~I_Q(X;X^\prime) <
2R-\epsilon\}}[\calG_n(Q_{XX^\prime Y})\cap\calE^c(Q_{XX^\prime})].
\end{equation}
We first show that $\mbox{Pr}\{\calG_n\}\to 1$ as $n\to\infty$.
We have already shown that $\mbox{Pr}\{\calE(Q_{XX^\prime})\}\le
e^{-n[2R-I_Q(X;X^\prime)]}\le e^{-n\epsilon}$. As for $\calG_n(Q_{XX^\prime
Y})$, we have
the following consideration. By the Chebychev inequality
\begin{eqnarray}
& &\mbox{Pr}\{[\calG_n(Q_{XX^\prime Y})]^c\}\nonumber\\
&\le&\mbox{Pr}\left[\sum_m\sum_{m^\prime\ne
m}\calI\{(\bx_m\bx_{m^\prime})\in\calT(Q_{XX^\prime})\}\cdot
\sum_{\by\in\calT(Q_{Y|XX^\prime}|\bx_m\bx_{m^\prime})}\calI\{Z_{mm^\prime}(\by)>
e^{n[\alpha(R+2\epsilon,Q_Y)+\epsilon]}\}>\right.\nonumber\\
& &\left.\exp\{n[2R-I_Q(X;X^\prime)-3\epsilon/2]\}\cdot
|\calT(Q_{Y|XX^\prime}|\bx_m\bx_{m^\prime})|\right]\nonumber\\
&\le&\frac{\bE\left\{\sum_m\sum_{m^\prime\ne
m}\calI\{(\bx_m\bx_{m^\prime})\in\calT(Q_{XX^\prime})\}\cdot
\sum_{\by\in\calT(Q_{Y|XX^\prime}|\bx_m\bx_{m^\prime})}\calI\{Z_{mm^\prime}(\by)>
e^{n[\alpha(R+2\epsilon,Q_Y)+\epsilon]}\}\right\}}{
\exp\{n[2R-I_Q(X;X^\prime)-3\epsilon/2]\}\cdot
|\calT(Q_{Y|XX^\prime}|\bx_m\bx_{m^\prime})|}\nonumber\\
&\le&\frac{e^{2nR}|\calT(Q_{Y|XX^\prime}|\bx_m\bx_{m^\prime})|\cdot
\mbox{Pr}\{(\bX_m,\bX_{m^\prime})\in\calT(Q_{XX^\prime}),~Z_{mm^\prime}(\by)\ge
e^{n[\alpha(R+2\epsilon,Q_Y)+\epsilon]}\}}{\exp\{n[2R-I_Q(X;X^\prime)-3\epsilon/2]\}\cdot
|\calT(Q_{Y|XX^\prime}|\bx_m\bx_{m^\prime})|}\nonumber\\
&=&\frac{
\mbox{Pr}\{(\bX_m,\bX_{m^\prime})\in\calT(Q_{XX^\prime})\}\cdot\mbox{Pr}\{Z_{mm^\prime}(\by)\ge
e^{n[\alpha(R+2\epsilon,Q_Y)+\epsilon]}\}}{\exp\{-n[I_Q(X;X^\prime)+3\epsilon/2]\}}\nonumber\\
&\exe&\frac{
e^{-nI_Q(X;X^\prime)}\cdot
\mbox{Pr}\{Z_{mm^\prime}(\by)\ge
e^{n[\alpha(R+2\epsilon,Q_Y)+\epsilon]}\}}{\exp\{-n[I_Q(X;X^\prime)+3\epsilon/2]\}}\nonumber\\
&=&e^{3n\epsilon/2}\cdot \mbox{Pr}\{Z_{mm^\prime}(\by)\ge
e^{n[\alpha(R+2\epsilon,Q_Y)+\epsilon]}\}.
\end{eqnarray}
But
\begin{eqnarray}
& &\mbox{Pr}\{Z_{mm^\prime}(\by)\ge
e^{n[\alpha(R+2\epsilon,Q_Y)+\epsilon]}\}\nonumber\\
&=&\mbox{Pr}\left\{\sum_{Q_{X|Y}}N(Q_{XY})e^{ng(Q_{XY})}\ge
e^{n[\alpha(R+2\epsilon,Q_Y)+\epsilon]}\right\}\nonumber\\
&\exe&\max_{Q_{X|Y}}\mbox{Pr}\left\{N(Q_{XY})\ge
\exp\{n[\alpha(R+2\epsilon,Q_Y)+\epsilon-g(Q_{XY})]\}\right\}\nonumber\\
&\exe& e^{-nE},
\end{eqnarray}
where $N(Q_{XY})$ is the number of codewords other than $\bx_m$ and
$\bx_{m^\prime}$ that, together with $\by$, fall in $\calT(Q_{XY})$, which is
a binomial random variables with $e^{nR}-2$ trials and success rate of
exponential order $e^{-nI_Q(X;Y)}$, and so,
\begin{eqnarray}
E&=&\min_{Q_{X|Y}}\left\{\begin{array}{ll}
[I_Q(X;Y)-R]_+ & g(Q_{XY})+[R-I_Q(X;Y)]_+\ge
\alpha(R+2\epsilon,Q_Y)+\epsilon\\
\infty & g(Q_{XY})+[R-I_Q(X;Y)]_+<
\alpha(R+2\epsilon,Q_Y)+\epsilon\end{array}\right.\nonumber\\
&=&\inf_{\{Q_{X|Y}:~g(Q_{XY})+[R-I_Q(X;Y)]_+\ge
\alpha(R+2\epsilon,Q_Y)+\epsilon\}}[I_Q(X;Y)-R]_+.
\end{eqnarray}
Now, by definition of the function $\alpha(R,Q_Y)$, the set
$\{Q_{X|Y}:~g(Q_{XY})+[R-I_Q(X;Y)]_+\ge
\alpha(R+2\epsilon,Q_Y)+\epsilon\}$ is a subset of $\{Q_{X|Y}:~I_Q(X;Y)\ge
R+2\epsilon\}$. Thus,
\begin{equation}
E\ge \inf_{\{Q_{X|Y}:~I_Q(X;Y)\ge R+2\epsilon\}}[I_Q(X;Y)-R]_+=2\epsilon,
\end{equation}
and so, $\mbox{Pr}\{Z_{mm^\prime}(\by)\ge
e^{n[\alpha(R+2\epsilon,Q_Y)+\epsilon]}\}\lexe
e^{-2n\epsilon}$, which leads to
\begin{equation}
\mbox{Pr}\{[\calG_n(Q_{XX^\prime Y})]^c\}\lexe e^{3n\epsilon/2}\cdot
e^{-2n\epsilon}=e^{-n\epsilon/2}.
\end{equation}
Since the number of types is merely polynomial, it follows that
$\mbox{Pr}\{\calG_n\}\to 1$.
Now, for a given $\calC_n\in\calG(Q_{XX^\prime Y})$, let us define the set
\begin{equation}
\calF(\calC_n,Q_{XX^\prime Y})=\{(m,m^\prime,\by):~Z_{mm^\prime}(\by)
\le\exp\{n[\alpha(R+2\epsilon,Q_Y)+\epsilon]\}\},
\end{equation}
and
\begin{equation}
\calF(\calC_n,Q_{XX^\prime
Y}|m,m^\prime)=\{\by:~(m,m^\prime,\by)\in\calF(\calC_n,Q_{XX^\prime Y})\}.
\end{equation}
Then, by definition, for $\calC_n\in\calG_n(Q_{XX^\prime Y})$,
\begin{eqnarray}
& &\sum_{m,m^\prime}\calI\{(\bx_m\bx_{m^\prime})\in\calT(Q_{XX^\prime})\}\cdot\frac{
|\calT(Q_{Y|XX^\prime}|\bx_m\bx_{m^\prime})\cap\calF(\calC_n,Q_{XX^\prime
Y}|m,m^\prime)|}{|\calT(Q_{Y|XX^\prime}|\bx_m\bx_{m^\prime})|}\nonumber\\
&\ge&
\exp\{n[2R-I_Q(X;X^\prime)-3\epsilon/2\},
\end{eqnarray}
where we have used the fact that
$|\calT(Q_{Y|XX^\prime}|\bx_m\bx_{m^\prime})|$ is the same for all
$(\bx_m\bx_{m^\prime})\in\calT(Q_{XX^\prime})$.
Putting all this together, we now have:
\begin{eqnarray}
& &\bE\left\{\left[P_{\mbox{\tiny e}}(\calC_n)\right]^{1/\rho}\right\}\nonumber\\
&=&\bE\left[\frac{1}{M}\sum_m\sum_{m^\prime\ne
m}\sum_{\by}W(\by|\bx_m)\cdot\frac{\exp\{ng(\hP_{\bx_{m^\prime}\by})\}}
{\exp\{ng(\hP_{\bx_m\by})\}+\exp\{ng(\hP_{\bx_{m^\prime}\by})\}+
Z_{mm^\prime}(\by)}\right]^{1/\rho}\nonumber\\
&=&\sum_{\calC_n}P(\calC_n)
\left[\frac{1}{M}\sum_m\sum_{m^\prime\ne
m}\sum_{\by}W(\by|\bx_m)\cdot\frac{\exp\{ng(\hP_{\bx_{m^\prime}\by})\}}
{\exp\{ng(\hP_{\bx_m\by})\}+\exp\{ng(\hP_{\bx_{m^\prime}\by})\}+
Z_{mm^\prime}(\by)}\right]^{1/\rho}\nonumber\\
&\ge&
\sum_{\calC_n\in\calG_n}P(\calC_n)
\left[\frac{1}{M}\sum_{\{Q_{XX^\prime}:~I_Q(X;X^\prime) < 2R-\epsilon\}}
\sum_m\sum_{m^\prime\ne
m}\calI\{(\bx_m\bx_{m^\prime})\in\calT(Q_{XX^\prime})\}\times\right.\nonumber\\
& &\left.\sum_{Q_{Y|XX^\prime}}
\sum_{\by\in
\calT(Q_{Y|XX^\prime}|\bx_m\bx_{m^\prime})\cap\calF(\calC_n,Q_{XX^\prime
Y}|m,m^\prime)}W(\by|\bx_m)\times\right.\nonumber\\
& &\left.\frac{\exp\{ng(\hP_{\bx_{m^\prime}\by})\}}
{\exp\{ng(\hP_{\bx_m\by})\}+\exp\{ng(\hP_{\bx_{m^\prime}\by})\}+
Z_{mm^\prime}(\by)}\right]^{1/\rho}\nonumber\\
&\ge&
\sum_{\calC_n\in\calG_n}P(\calC_n)
\left[\frac{1}{M}\sum_{\{Q_{XX^\prime}:~I_Q(X;X^\prime) < 2R-\epsilon\}}
\sum_m\sum_{m^\prime\ne
m}\calI\{(\bx_m\bx_{m^\prime})\in\calT(Q_{XX^\prime})\}\times\right.\nonumber\\
& &\left.\sum_{Q_{Y|XX^\prime}}
\sum_{\by\in
\calT(Q_{Y|XX^\prime}|\bx_m\bx_{m^\prime})\cap\calF(\calC_n,Q_{XX^\prime
Y}|m,m^\prime)}W(\by|\bx_m)\times\right.\nonumber\\
& &\left.\frac{\exp\{ng(\hP_{\bx_{m^\prime}\by})\}}
{\exp\{ng(\hP_{\bx_m\by})\}+\exp\{ng(\hP_{\bx_{m^\prime}\by})\}+
\exp\{n[\alpha(R+2\epsilon,Q_Y)+\epsilon]\}}\right]^{1/\rho}\nonumber\\
&\ge&
\sum_{\calC_n\in\calG_n}P(\calC_n)
\left[\frac{1}{M}\sum_{\{Q_{XX^\prime}:~I_Q(X;X^\prime) < 2R-\epsilon\}}
\sum_m\sum_{m^\prime\ne
m}\calI\{(\bx_m\bx_{m^\prime})\in\calT(Q_{XX^\prime})\}\times\right.\nonumber\\
& &\left.\sum_{Q_{Y|XX^\prime}}
\frac{|\calT(Q_{Y|XX^\prime}|\bx_m\bx_{m^\prime})\cap\calF(\calC_n,Q_{XX^\prime
Y}|m,m^\prime)|}{|\calT(Q_{Y|XX^\prime}|\bx_m\bx_{m^\prime})|}\cdot
|\calT(Q_{Y|XX^\prime}|\bx_m\bx_{m^\prime})|\cdot
W(\by|\bx_m)\times\right.\nonumber\\
& &\left.\frac{\exp\{ng(\hP_{\bx_{m^\prime}\by})\}}
{\exp\{ng(\hP_{\bx_m\by})\}+\exp\{ng(\hP_{\bx_{m^\prime}\by})\}+
\exp\{n[\alpha(R+2\epsilon,Q_Y)+\epsilon]\}}\right]^{1/\rho}\nonumber\\
&\ge&
\sum_{\calC_n\in\calG_n}P(\calC_n)
\left[\frac{1}{M}\sum_{\{Q_{XX^\prime Y}:~I_Q(X;X^\prime) < 2R-\epsilon\}}
\sum_m\sum_{m^\prime\ne
m}\calI\{(\bx_m\bx_{m^\prime})\in\calT(Q_{XX^\prime})\}\times\right.\nonumber\\
& &\left.
\frac{|\calT(Q_{Y|XX^\prime}|\bx_m\bx_{m^\prime})\cap\calF(\calC_n,Q_{XX^\prime
Y}|m,m^\prime)|}{|\calT(Q_{Y|XX^\prime}|\bx_m\bx_{m^\prime})|}\cdot
|\calT(Q_{Y|XX^\prime}|\bx_m\bx_{m^\prime})|\cdot
W(\by|\bx_m)\times\right.\nonumber\\
& &\left.\frac{\exp\{ng(\hP_{\bx_{m^\prime}\by})\}}
{\exp\{ng(\hP_{\bx_m\by})\}+\exp\{ng(\hP_{\bx_{m^\prime}\by})\}+
\exp\{n[\alpha(R+2\epsilon,Q_Y)+\epsilon]\}}\right]^{1/\rho}\nonumber\\
&\ge&
P(\calC_n)
\left[\sum_{\{Q_{XX^\prime Y}:~I_Q(X;X^\prime) < 2R-\epsilon\}}
\exp\{n[R-I_Q(X;X^\prime)-3\epsilon/2]\}\times\right.\nonumber\\
& &\left.\exp\{-n[D(Q_{Y|X}\|W|Q_X)+I_Q(X^\prime;Y|X)+\right.\nonumber\\
& &\left.[\max\{g(Q_{XY}),\alpha(R+2\epsilon,Q_Y)+\epsilon\}-g(Q_{X^\prime Y})]_+\}
\right]^{1/\rho}\nonumber\\
&\exe&\exp\{-n[E_{\mbox{\tiny trc}}^g(R,Q_X)+O(\epsilon)]/\rho\},
\end{eqnarray}
which after raising to the power of $\rho$, gives the desired result, and
completes the proof due to the arbitrariness of $\epsilon$.

\section{TRC Error Exponents in More General Settings}

In this section, we demonstrate that the same analysis technique is applicable
to other, more general scenarios of coded communication systems. We briefly outline
the analysis and the resulting TRC error exponents
in two examples of such scenarios. The first is list decoding where the
list size $L$ is fixed, independently of $n$. For simplicity, we take $L=2$,
but the extension to general $L$ will be straightforward. The second is
decoding with an erasure/list option in the framework of Forney
\cite{Forney68}, where we analyze the exponential rate of the undetected error
of the TRC. In both examples, we continue to consider 
the ensemble of fixed composition codes of type $Q_X$, and we 
allow a general decoding metric $g$, as
before.

\subsection{List Decoding}

Consider a list decoder of list--size $L$ and a deterministic decoder with decoding metric $g$. 
Such a decoder outputs the list of the $L$ messages with the highest scores,
$g(\hP_{\bx_m\by})$. A list error is the event that the correct codeword is
not in the list.
For simplicity, we take $L=2$, but the treatment for a
general $L$ will be self--evident.

As we observed earlier, in the limit of deterministic decoding
($\beta\to\infty$), for the vast majority of codes, the highest score of an
incorrect codeword is typically no larger than
$a(R,Q_Y)=\max\{g(Q_{XY}):~I_Q(X;Y)\le R,~(Q_Y\odot Q_{X|Y})_X=Q_X\}$ whenever $\by\in\calT(Q_Y)$.
Let us define
\begin{equation}
\Lambda_{\mbox{\tiny
L}}(Q_{XX^\prime\tilde{X}})\dfn\inf_{Q_{Y|XX^\prime\tilde{X}}\in\calQ}\{D(Q_{Y|X}\|W|Q_X)+
I_Q(X^\prime;Y|X)\},
\end{equation}
where
\begin{equation}
\calQ=\{Q_{Y|XX^\prime\tilde{X}}:~g(Q_{X^\prime Y})\ge
g(Q_{\tilde{X}Y})\ge\max\{g(Q_{XY}),a(R,Q_Y)\}\}.
\end{equation}
Then, the TRC list error exponent is given by
\begin{equation}
E_{\mbox{\tiny trcl}}^g
(R,Q_X)=\inf_{Q_{X^\prime\tilde{X}|X}\in\calS(R)}\{\Lambda_{\mbox{\tiny
L}}(Q_{XX^\prime\tilde{X}})+I_Q(X;X^\prime;\tilde{X})-2R\},
\end{equation}
where $I_Q(X;X^\prime;\tilde{X})$ is the multi--information, defined as
\begin{equation}
I_Q(X;X^\prime;\tilde{X})=H_Q(X)+H_Q(X^\prime)+H_Q(\tilde{X})-H_Q(X,X^\prime,\tilde{X})
\end{equation}
and
\begin{eqnarray}
\calS(R)&=&\{Q_{X^\prime\tilde{X}|X}:~
\max\{I_Q(X;X^\prime),
I_Q(X^\prime;\tilde{X}),I_Q(X;\tilde{X})\}< 2R,~\nonumber\\
& &I_Q(X;X^\prime;\tilde{X}) < 3R,~Q_{X^\prime}=Q_{\tilde{X}}=Q_X\}.
\end{eqnarray}

We next provide a brief outline of the derivation, which is largely quite a
simple extension of the first part of the proof of Theorem \ref{main}, but
with a few twists.
\begin{eqnarray}
\bE\left\{\left[P_{\mbox{\tiny e}}^{\mbox{\tiny
list}}(\calC_n)\right]^{1/\rho}\right\}
&\lexe&\bE\left\{\left[\frac{1}{M}\sum_m\sum_{m^\prime\ne m}\sum_{\tilde{m}\ne
m,m^\prime}\sum_{\by}W(\by|\bx_m)\times\right.\right.\nonumber\\
& &\left.\left.\calI\left\{g(\hP_{\bx_{m^\prime}\by})\ge
g(\hP_{\bx_{\tilde{m}}\by})\ge\max\{g(\hP_{\bx_m\by}),\max_{\hat{m}\ne
m,m^\prime}g(\hP_{\bx_{\hat{m}}\by})\}\right\}\right]^{1/\rho}\right\}\nonumber\\
&\lexe&\bE\left\{\left[\frac{1}{M}\sum_m\sum_{m^\prime\ne m}\sum_{\tilde{m}\ne
m,m^\prime}\sum_{\by}W(\by|\bx_m)\times\right.\right.\nonumber\\
& &\left.\left.\calI\left\{g(\hP_{\bx_{m^\prime}\by})\ge
g(\hP_{\bx_{\tilde{m}}\by})\ge\max\{g(\hP_{\bx_m\by}),a(R-\epsilon,\hP_{\by})\}
\right\}\right]^{1/\rho}\right\}\nonumber\\
&\lexe&\bE\left\{\left[\frac{1}{M}\sum_{Q_{XX^\prime\tilde{X}}}
N(Q_{XX^\prime\tilde{X}})\exp\left\{-n\Lambda_{\mbox{\tiny
L}}(Q_{XX^\prime\tilde{X}})\right\}
\right]^{1/\rho}\right\}\nonumber\\
&\lexe&M^{-1/\rho}\sum_{Q_{XX^\prime\tilde{X}}}
\bE\{[N(Q_{XX^\prime\tilde{X}})]^{1/\rho}\cdot\exp\left\{-n\Lambda_{\mbox{\tiny
L}}(Q_{XX^\prime\tilde{X}})/\rho\right\},
\end{eqnarray}
where $N(Q_{XX^\prime\tilde{X}})$ is the number of codeword triplets,
$\{\bx_m,\bx_{m^\prime},\bx_{\tilde{m}}\}$, ($m$, $m^\prime$ and $\tilde{m}$
-- all distinct) that fall within $\calT(Q_{XX^\prime\tilde{X}})$.
Now,
\begin{equation}
\label{series}
\bE\{[N(Q_{XX^\prime\tilde{X}})]^{1/\rho}\}=\sum_{k\ge
1}\mbox{Pr}\{N(Q_{XX^\prime})=k\}\cdot
\bE\{[N(Q_{XX^\prime\tilde{X}})]^{1/\rho}\bigg|N(Q_{XX^\prime})=k\}.
\end{equation}
Consider first the case $I_Q(X;X^\prime)> 2R$. Then the sum on the r.h.s.\ of
(\ref{series}) is
dominated by $k\exe 1$, and so,
\begin{equation}
\bE\{[N(Q_{XX^\prime\tilde{X}})]^{1/\rho}\}\exe
e^{n[2R-I_Q(X;X^\prime)]}\cdot\left\{\begin{array}{ll}
e^{n[R-I_Q(X,X^\prime;\tilde{X})]/\rho} & I_Q(X,X^\prime;\tilde{X}) < R\\
e^{n[R-I_Q(X,X^\prime;\tilde{X})]} & 
I_Q(X,X^\prime;\tilde{X}) > R\end{array}\right.
\end{equation}
which yields
\begin{equation}
\lim_{\rho\to\infty}\left(\bE\{[N(Q_{XX^\prime\tilde{X}})]^{1/\rho}\}\right)^\rho\lexe
\left\{\begin{array}{ll}
0 & I_Q(X,X^\prime;\tilde{X}) < R\\
0 & I_Q(X,X^\prime;\tilde{X})>R
\end{array}\right.=0.
\end{equation}
Similarly, $\bE\{[N(Q_{XX^\prime\tilde{X}})]^{1/\rho}\}\exe 0$ as well when
$I_Q(X^\prime;\tilde{X})> 2R$ or
$I_Q(X;\tilde{X})> 2R$.
In the case $I_Q(X;X^\prime)< 2R$, 
the sum on the r.h.s.\ of (\ref{series}) is dominated by $k\exe
e^{n[2R-I_Q(X;X^\prime)]}$, and so,
\begin{eqnarray}
\bE\{[N(Q_{XX^\prime\tilde{X}})]^{1/\rho}\}&\lexe&
e^{n[2R-I_Q(X;X^\prime)]s/\rho}\cdot\left(\bE[N(Q_{\tilde{X}|XX^\prime})]^{1/s}\right)^{s/\rho}\nonumber\\
&\exe&e^{n[2R-I_Q(X;X^\prime)]s/\rho}\cdot\left\{\begin{array}{ll}
e^{n[R-I_Q(X,X^\prime;\tilde{X})]/\rho} & I_Q(X,X^\prime;\tilde{X}) < R\\
e^{n[R-I_Q(X,X^\prime;\tilde{X})]s/\rho} & 
I_Q(X,X^\prime;\tilde{X})> R\end{array}\right.\nonumber\\
&=&\left\{\begin{array}{ll}
e^{n[s(2R-I_Q(X;X^\prime))+R-I_Q(X,X^\prime;\tilde{X})]/\rho} & 
I_Q(X,X^\prime;\tilde{X}) < R\\
e^{n[3R-I_Q(X;X^\prime;\tilde{X})]s/\rho} & 
I_Q(X,X^\prime;\tilde{X})> R\end{array}\right.
\end{eqnarray}
and after optimizing over $s$,
\begin{eqnarray}
\bE\{[N(Q_{XX^\prime\tilde{X}})]^{1/\rho}\}&\lexe&\left\{\begin{array}{ll}
e^{n[3R-I_Q(X;X^\prime;\tilde{X})]/\rho}  & I_Q(X,X^\prime;\tilde{X}) < R\\
e^{n[3R-I_Q(X;X^\prime;\tilde{X})]/\rho}  & I_Q(X,X^\prime;\tilde{X})> R,~
I_Q(X;X^\prime;\tilde{X})< 3R\\
e^{n[3R-I_Q(X;X^\prime;\tilde{X})]}  & I_Q(X,X^\prime;\tilde{X})> R,~
I_Q(X;X^\prime;\tilde{X}) > 3R\end{array}\right.
\end{eqnarray}
and so,
\begin{eqnarray}
\lim_{\rho\to\infty}\left(\bE\{[N(Q_{XX^\prime\tilde{X}})]^{1/\rho}\}\right)^\rho&\lexe&
\left\{\begin{array}{ll}
e^{n[3R-I_Q(X;X^\prime;\tilde{X})]}  & I_Q(X,X^\prime;\tilde{X}) < R\\
e^{n[3R-I_Q(X;X^\prime;\tilde{X})]}  & I_Q(X,X^\prime;\tilde{X}) > R,~
I_Q(X;X^\prime;\tilde{X}) < 3R\\
0 & R < I_Q(X,X^\prime;\tilde{X}),~
I_Q(X;X^\prime;\tilde{X})> 3R\end{array}\right.\nonumber\\
&=&\left\{\begin{array}{ll}
e^{n[3R-I_Q(X;X^\prime;\tilde{X})]}  & 
I_Q(X;X^\prime;\tilde{X}) < 3R\\
0 & 
I_Q(X;X^\prime;\tilde{X}) > 3R\end{array}\right.
\end{eqnarray}
A similar derivation applies to the two other combinations of two out the
three random variables, $X$, $X^\prime$ and $\tilde{X}$.
In summary,
\begin{equation}
\lim_{\rho\to\infty}\left(\bE
\{[N(Q_{XX^\prime\tilde{X}})]^{1/\rho}\}\right)^\rho\nonumber\\
\lexe
\left\{\begin{array}{ll}
e^{n[3R-I_Q(X;X^\prime;\tilde{X})]}  & Q_{X^\prime\tilde{X}|X}\in\calS(R)\\
0 & \mbox{elsewhere}\end{array}\right.
\end{equation}
and the desired result follows similarly as before.

\subsection{Decoding with an Erasure/List Option}

Consider the following generalized version of Forney's erasure/list decoder
\cite{Forney68}, which for a
given parameter $T$, decides in favor of message $m$ whenever
\begin{equation}
\frac{\exp\{ng(\hP_{\bx_m\by})\}}{\sum_{m^\prime\ne
m}\exp\{ng(\hP_{\bx_{m^\prime}\by})\}}\ge e^{nT},
\end{equation}
and erases if no message $m$ satisfies this inequality.

We now define
\begin{eqnarray}
& &\Lambda(Q_{XX^\prime},R,T)\nonumber\\
&=&\min_{\{Q_{Y|XX^\prime}:~g(Q_{X^\prime
Y})-\max\{g(Q_{XY}),\alpha(R,Q_Y)\}\ge
T\}}\left\{D(Q_{Y|X}\|W|Q_X)+I_Q(X^\prime;Y|X)\right\}.
\end{eqnarray}
and then we argue that TRC undetected error exponent is given by
\begin{equation}
E_{\mbox{\tiny trc-ue}}^g(R,T,Q_X)=
\min_{\{Q_{XX^\prime}:~I_Q(X;X^\prime)\le
2R,~Q_{X^\prime}=Q_X\}}[\Lambda(Q_{XX^\prime},R,T)+I_Q(X;X^\prime)-R].
\end{equation}

The outline is for the derivation is as follows.
For a given code $\calC_n$, the probability of undetected error is given by
\begin{eqnarray}
P_{\mbox{\tiny ue}}(\calC_n)&=&\frac{1}{M}\sum_{m=0}^{M-1}
\mbox{Pr}\bigcup_{m^\prime\ne
m}\left\{\frac{\exp\{ng(\hP_{\bx_{m^\prime}\by})\}}{\sum_{\tilde{m}\ne
m^\prime}\exp\{ng(\hP_{\bx_{\tilde{m}}\by})\}}\ge e^{nT}\right\}\nonumber\\
&=&\frac{1}{M}\sum_{m=0}^{M-1}
\sum_{m^\prime\ne
m}\mbox{Pr}\left\{\frac{\exp\{ng(\hP_{\bx_{m^\prime}\by})\}}{\sum_{\tilde{m}\ne
m^\prime}\exp\{ng(\hP_{\bx_{\tilde{m}}\by})\}}\ge e^{nT}\right\}\nonumber\\
&=&\frac{1}{M}\sum_{m=0}^{M-1}
\sum_{m^\prime\ne
m}\mbox{Pr}\left\{\frac{\exp\{ng(\hP_{\bx_{m^\prime}\by})\}}
{\exp\{ng(\hP_{\bx_m\by})\}+\sum_{\tilde{m}\ne
m,m^\prime}\exp\{ng(\hP_{\bx_{\tilde{m}}\by})\}}\ge e^{nT}\right\}\nonumber\\
&\dfn&\frac{1}{M}\sum_{m=0}^{M-1}
\sum_{m^\prime\ne
m}\sum_{\by}W(\by|\bx_m)\cdot\calI\left\{\frac{\exp\{ng(\hP_{\bx_{m^\prime}\by})\}}
{\exp\{ng(\hP_{\bx_m\by})\}+
Z_{mm^\prime}(\by)}\ge e^{nT}\right\}.
\end{eqnarray}
Thus, using the same considerations as before,
\begin{eqnarray}
& &\bE\left\{\left[P_{\mbox{\tiny
ue}}(\calC_n)\right]^{1/\rho}\right\}\nonumber\\
&=&\bE\left\{\left[\frac{1}{M}\sum_{m=0}^{M-1}
\sum_{m^\prime\ne
m}\sum_{\by}W(\by|\bx_m)\cdot\calI\left\{\frac{\exp\{ng(\hP_{\bx_{m^\prime}\by})\}}
{\exp\{ng(\hP_{\bx_m\by})\}+
Z_{mm^\prime}(\by)}\ge e^{nT}\right\}\right]^{1/\rho}\right\}\nonumber\\
&\lexe&
\bE\left\{\left[\frac{1}{M}\sum_{m=0}^{M-1}
\sum_{m^\prime\ne
m}\sum_{\by}W(\by|\bx_m)\times\right.\right.\nonumber\\
& &\left.\left.\calI\left\{\frac{\exp\{ng(\hP_{\bx_{m^\prime}\by})\}}
{\exp\{ng(\hP_{\bx_m\by})\}+\exp\{n\alpha(R-\epsilon,\hP_{\by})\}}\ge
e^{nT}\right\}\right]^{1/\rho}\right\}\nonumber\\
&\exe&
\bE\left\{\left[\frac{1}{M}\sum_{m=0}^{M-1}
\sum_{m^\prime\ne
m}\sum_{\by}W(\by|\bx_m)\times\right.\right.\nonumber\\
& &\left.\left.\calI\left\{g(\hP_{\bx_{m^\prime}\by})-
\max\{g(\hP_{\bx_m\by}),\alpha(R-\epsilon,\hP_{\by})\ge T
\right\}\right]^{1/\rho}\right\}.
\end{eqnarray}
By the method of types \cite{CK11}, the inner--most sum is of 
the exponential order of $\exp\{-n\Lambda(\hP_{\bx_m\bx_{m^\prime}},R,T)\}$, and
so,
\begin{eqnarray}
\bE\left\{\left[P_{\mbox{\tiny
ue}}(\calC_n)\right]^{1/\rho}\right\}&\lexe&
\bE\left\{\left[\frac{1}{M}\sum_{m=0}^{M-1}
\sum_{m^\prime\ne
m}\exp\{-n\Lambda(\hP_{\bx_m\bx_{m^\prime}},R,T)\}\right]^{1/\rho}\right\}\nonumber\\
&=&M^{-1/\rho}\bE\left\{\left[\sum_{Q_{XX^\prime}}N(Q_{XX^\prime})
\exp\{-n\Lambda(Q_{XX^\prime},R,T)\}\right]^{1/\rho}\right\}\nonumber\\
&\lexe&M^{-1/\rho}\sum_{Q_{XX^\prime}}\bE\left\{\left[N(Q_{XX^\prime})\right]^{1/\rho}
\right\}\cdot\exp\{-n\Lambda(Q_{XX^\prime},R,T)/\rho\},
\end{eqnarray}
which yields, similarly as before,
\begin{eqnarray}
& &\lim_{\rho\to\infty}\left(\bE\left\{\left[P_{\mbox{\tiny
ue}}(\calC_n)\right]^{1/\rho}\right\}\right)^\rho\nonumber\\
&\exe&
\exp\left\{-n\min_{\{Q_{XX^\prime}:~I_Q(X;X^\prime)\le
2R,~Q_{X^\prime}=Q_X\}}[\Lambda(Q_{XX^\prime},R,T)+I_Q(X;X^\prime)-R]\right\}\nonumber\\
&=&\exp\{-nE_{\mbox{\tiny trc-ue}}^g(R,T,Q_X)\}.
\end{eqnarray}

\section*{Acknowledgement}

Interesting discussions with Anelia Somekh--Baruch, in the early stages of this
work, are acknowledged with thanks.


\end{document}